\documentclass{appolb}
\usepackage{graphicx}
\usepackage{amsmath,amssymb}
\usepackage{url}
\usepackage{epsfig}
\usepackage[T1]{fontenc}
\usepackage[latin9]{inputenc}
\usepackage{multirow}

\newcommand{\tb}{\bar t}

\newcommand{\ttbh}{ t \tb H}
\newcommand{\ttbz}{ t \tb Z}
\newcommand{\ttbw}{ t \tb W}
\newcommand{\ttb}{ t \tb B}

\newcommand{\als}{\alpha_{\rm s}}
\newcommand{\shat}{\hat s}

\newcommand{\muf}{\mu_{\rm F}}
\newcommand{\mur}{\mu_{\rm R}}
\newcommand{\mufo}{\mu_{{\rm F},0}}
\newcommand{\muro}{\mu_{{\rm R},0}}
\newcommand{\sigh}{\hat \sigma}

\newcommand{\nn}{\nonumber}

\newcommand{\tosv}{{\scriptscriptstyle \to}}

\def\beq{\begin{equation}}
\def\eeq{\end{equation}}
\def\bear{\begin{eqnarray}}
\def\eear{\end{eqnarray}}

\def\bet34{\beta_{kl}}

\begin{document}
\title{Top Precision for  Associated Top-Pair Production Processes at the LHC
\thanks{Presented by A. Kulesza at XXVI Cracow EPIPHANY Conference on LHC Physics: Standard Model and Beyond}%
}
\author{Anna Kulesza
\address{Institute of Theoretical Physics, WWU M\"unster, D-48149 M\"unster, Germany}\\[1em]
Leszek Motyka
\address{Institute of Theoretical Physics, Jagiellonian University, 30-348 Krak\'ow, Poland}\\[1em]
{Daniel Schwartl\"ander
}
\address{Institute of Theoretical Physics, WWU M\"unster, D-48149 M\"unster, Germany}\\[1em]
{ Tomasz Stebel
}
\address{Institute of Theoretical Physics, Jagiellonian University, 30-348 Krak\'ow, Poland}\\[1em]
{ Vincent Theeuwes
}
\address{Institute for Theoretical Physics, Georg-August-University G\"ottingen,  37077 G\"ottingen, Germany}
}
\maketitle
\begin{abstract}
The studies of the associated production processes of a top-quark pair with a colour-singlet boson, e.g. Higgs, W or Z, are among the highest priorities of the LHC programme. Correspondingly, improvements in precision of theoretical predictions for these processes are of central importance. In this talk, we review our latest results on resummation of soft gluon corrections. The resummation is carried out using the direct QCD Mellin space technique in three-particle invariant mass kinematics. We discuss the impact of the soft gluon corrections on predictions for total cross sections and  differential distributions.
\end{abstract}
  
\section{Introduction}
The measurements \cite{Sirunyan:2018hoz}--\cite{CMS:2019too}
%\cite{Sirunyan:2018hoz, Aaboud:2018urx, Chatrchyan:2013qca,Khachatryan:2014ewa,Aad:2015eua,Khachatryan:2015sha,Aaboud:2016xve,Sirunyan:2017uzs,  Aaboud:2019njj,CMS:2019too} 
of associated production of a heavy boson ($H$,$W$, $Z$) with a top-antitop quark pair provide an important test for the Standard Model at the Large Hadron Collider (LHC). 
These are the key processes to experimentally determine the top quark couplings. In particular, the associated $t \bar t  H$ production  directly probes the top Yukawa coupling without making any assumptions on its nature. Moreover they are relevant in searches for new physics due to both being directly sensitive to it and providing an important background. The $t \bar t W$,  $t \bar t Z$ processes also play an important role as a background for the associated Higgs boson production process $pp \rightarrow t \bar t H$. Thus it is necessary to know the theoretical predictions for $pp \rightarrow t \bar t B$, $B=H,W^\pm,Z$ with high accuracy, especially in the light of ever improving precision of cross section measurements. For example, the very recent measurement of the $\ttbz$ cross section~\cite{CMS:2019too}  carries statitical and systematic errors of only 5-7\%.

Fixed order cross sections up to next-to-leading order in $\alpha_S$ are already known for some time both for the asociated Higgs boson~\cite{Beenakker:2001rj, Reina:2001sf} and $W$ and $Z$ boson production \cite{Lazopoulos:2008de,Lazopoulos:2007bv}. They were recalculated and matched to parton showers in \cite{Hirschi:2011pa, Frederix:2011zi, Garzelli:2011vp, Hartanto:2015uka, Kardos:2011na,Campbell:2012dh,Alwall:2014hca,Garzelli:2011is,Garzelli:2012bn}. Furthermore, QCD-EW NLO corrections are also known \cite{Frixione:2014qaa,Frixione:2015zaa,Yu:2014cka}.
For the $t \bar t H$ and $\ttbz$ processes, the NLO QCD~\cite{Denner:2015yca, Bevilacqua:2019cvp} and EW corrections~\cite{Denner:2016wet} to production with off-shell top quarks were also calculated. While NNLO calculations for this particular type of $2 \to 3$ processes are currently out of reach, a class of corrections beyond NLO from the emission of soft and/or collinear gluons can be taken into account with the help of resummation methods. Such methods allow to account for effects of soft gluon emission to all orders in perturbation theory. Two  common approaches to perform soft gluon resummation are either calculations directly in QCD or in an effective field theory, in this case soft-collinear effective theory (SCET). 

For the associated $\ttbh$  production, the first calculations of the resummed cross section at the next-to-leading logarithmic (NLL) acuracy, matched to the NLO result were presented in~\cite{Kulesza:2015vda}. The calculation relied on application of the traditional Mellin-space resummation formalism in the absolute threshold limit, i.e.\ in the limit of the partonic energy $\sqrt{\shat}$ approaching the production threshold $M=2 m_t + m_H$. Subsequently,  resummation of NLL corrections arising in the limit of $\sqrt{\shat}$ approaching the invariant mass threshold $Q$, with $Q^2= (p_t +p_{\bar t}+ p_H)^2$, was performed in~\cite{Kulesza:2016vnq} and later extended to the next-to-next-to-leading-logarithmic (NNLL) accuracy and applied to the $\ttbh$  production \cite{Kulesza:2017ukk}, as well as $\ttbz$ and $\ttbw$ production~\cite{Kulesza:2018tqz}. Apart for the total cross sections, also the distribution in the invariant mass $Q$~\cite{Kulesza:2017ukk,Kulesza:2018tqz, Kulesza:2020nfh}, the transverse momentum of the boson $B$, $p_T(B)$,   \cite{Kulesza:2019adl, Kulesza:2020nfh}, the invariant mass $m_{t \bar t}$ of the $t \bar t$ pair, transverse momentum of the top quark $p_T (t)$, the difference in rapidities between the top quark and the antitop quark $y(t) - y(\bar t)$, the difference in rapidities between the top quark  and the boson $y(t)-y(B)$, the difference in the azimuthal angle between the top quark and the antitop quark $\phi(t) - \phi(\bar t)$, and the difference in the azimuthal angle between the top quark and the boson $\phi(t) - \phi(B)$
\cite{Kulesza:2020nfh} were computed in the direct QCD approach. Some of these calculations \cite{Kulesza:2018tqz, Kulesza:2019adl, Kulesza:2020nfh} involved matching to complete NLO (QCD+EW) result, i.e. including all EW and QCD contributions up to NLO in the corresponding coupling constant. Calculations in the framework of the soft-collinear effective theory (SCET) for the $t\bar{t}H$ process led first to  obtaining approximate NNLO~\cite{Broggio:2015lya} and later full NNLL~\cite{Broggio:2016lfj} predictions. NNLL+NLO predictions have been obtained in SCET for  $pp \rightarrow t \bar t W$ \cite{Li:2014ula,Broggio:2016zgg} and for $pp \rightarrow t \bar t Z$ in~\cite{Broggio:2017kzi}. Results for a set of differential $t\bar t B$ distributions in the SCET approach can be found in \cite{Broggio:2019ewu}. 

Below we review results for threshold-resummed cross sections $pp \rightarrow t \bar t B$, $B=H,W,Z$  in the invariant mass kinematics, obtained using the Mellin-space approach at NNLL accuracy \cite{Kulesza:2017ukk,Kulesza:2018tqz,Kulesza:2019adl,Kulesza:2020nfh}, matched to the complete NLO (QCD+EW) predictions.

\section{Analytical description}
In the following  we treat the soft gluon corrections in the invariant mass kinematics, i.e we consider the limit $\hat \rho = Q^2/\hat s \rightarrow 1$ with $Q^2=(p_t + p_{\bar t} + p_{B})^2$.  
The logarithms resummed in the invariant mass threshold limit have the form
 $  \alpha_S^m \left(\frac{\log^n{(1-\hat\rho)}}{1-\hat\rho}\right)_{+} ,\ m \le 2n-1$
with the plus distribution $\int_0^1\text{d}x (f(x))_{+} = \int_0^1\text{d}x (f(x) - f(x_0))$.
The Mellin moments of the differential cross section $\text{d} \sigma_{ij \rightarrow t\bar t B}/{\text{d} Q^2}$ are taken with respect to the variable $\rho = Q^2/S$. At the partonic level this leads to
\begin{eqnarray}
&&\frac{\text{d} \tilde {\hat \sigma}_{ij \rightarrow t\bar t B}}{\text{d} Q^2} (N,Q^2,m_t,m_{W/Z},\mu_R^2,\mu_F^2)=\\ \nn
& &\int_0^1 \text{d} \hat \rho \hat \rho^{N-1} \frac{\text{d} \hat \sigma_{ij \rightarrow t\bar t B}}{\text{d} Q^2} (\hat \rho,Q^2,m_t,m_{W/Z},\mu_R^2,\mu_F^2)
\end{eqnarray}
for the Mellin moments for the process $ij \rightarrow t\bar t B$ with $i,j$ denoting two massless colored partons. In Mellin space the threshold limit $\hat \rho \rightarrow 1$ corresponds to the limit $N \rightarrow \infty$.
Since the process involves more than three colored partons, the resummed cross section is expressed in terms of  color matrices. In Mellin space the resummed partonic cross section has the form \cite{Contopanagos:1996nh,Kidonakis:1998nf}
\begin{equation}
 \frac{\text{d} \tilde {\hat \sigma}_{ij \rightarrow t\bar t B}}{\text{d} Q^2} = \text{Tr}[\mathbf{H}_{ij \rightarrow t\bar t B} \mathbf{S}_{ij \rightarrow t\bar t B}] \Delta_i \Delta_j,
\end{equation}
where $\mathbf{H}_{ij \rightarrow t\bar tW/Z}$ and $\mathbf{S}_{ij \rightarrow t\bar t B}$ are color matrices and the trace is taken in color space. We describe the evolution of color in the s-channel color basis, for which the basis vectors are
$
 c_{\mathbf{1}}=\delta_{a_i,a_j}\delta_{a_k,a_l} \quad c_{\mathbf{8}}=T^c_{a_i,a_j}T^c_{a_k,a_l}
$
for the $q \bar q$ initial state and 
$
 c_{\mathbf{1}}=\delta_{a_i,a_j}\delta_{a_k,a_l} \quad c_{\mathbf{8S}}=d^{c,a_i,a_j}T^c_{a_k,a_l} \quad c_{\mathbf{8A}}=f^{c,a_i,a_j}T^c_{a_k,a_l}
$
for the $gg$ initial state. This choice of color basis leads to a diagonal soft anomalous dimension matrix in the absolute threshold limit $(2m_t+m_{B})^2/{\hat s} \rightarrow 1$, which is a special case of the invariant mass threshold limit.
$\mathbf{H}_{ij \rightarrow t\bar t B}$ describes the hard scattering contributions projected on the color basis, while $\mathbf{S}_{ij \rightarrow t\bar t B}$ represents the soft wide angle emission. The (soft-)collinear logarithmic contributions from the initial state partons are taken into account by the functions $\Delta_i$ and $\Delta_j$. They  have been known for a long time \cite{Catani:1996yz,Bonciani:1998vc} and depend only on the emitting parton.

The soft function is given by a solution of the renormalization group equation~\cite{KS2,Czakon:2009zw}:
\bear
 \mathbf{S}_{ij\to klB}(N,Q^2,\muf^2, \mur^2)&=& \mathbf{\bar{U}}_{ij\tosv kl B}(N, Q^2,\muf^2, \mur^2 )\  \mathbf{\tilde S}_{ij\to klB}(\als(Q^2/{\bar N^2})) \nn \\
&& \mathbf{{U}}_{ij\tosv kl B}(N,Q^2,\muf^2, \mur^2), 
\label{eq:soft:evol}
\eear
where $\mathbf{\tilde S}_{ij\to klB}$ plays a role of a boundary condition. This soft matrix, as well as  the hard function $\mathbf{H}_{ij \rightarrow t\bar t B}$ can be calculated perturbatively:
$
\mathbf{\tilde S}_{ij\to klB}= \mathbf{\tilde S}^{\mathrm{(0)}}_{ij\to klB} + \frac{\als}{\pi}\mathbf{\tilde S}^{\mathrm{(1)}}_{ij\to klB} + \ldots,$ $
\mathbf{H}_{ ij\tosv kl B}= \mathbf{H}^{\mathrm{(0)}}_{ij\to klB} + \frac{\als}{\pi}\mathbf{H}^{\mathrm{(1)}}_{ij\to klB} +\ldots
\ $
At the NNLL accuracy knowledge of $\mathbf{\tilde S}^{\mathrm{(1)}}_{ij\to klB}$ and $\mathbf{H}^{\mathrm{(1)}}_{ij\to klB}$ is required whereas for NLL only leading terms  $\mathbf{H}^{\mathrm{(0)}}_{ij\to klB}$, $\mathbf{\tilde S}^{\mathrm{(0)}}_{ij\to klB}$ are needed.

The soft function evolution matrices  $\mathbf{{U}}_{ij\tosv kl B}$ are defined as a path-ordered exponents 
\beq
\mathbf{U}_{ij\to klB}\left(N, Q^2,\muf^2, \mur^2\right)=\mathrm{P}\exp\left[\int_{\muf}^{Q/\bar{N}}\frac{dq}{q}\pmb{\Gamma}_{ij\to klB}\left(\alpha_{\mathrm{s}}\left(q^{2}\right)\right)\right],
\label{eq:soft:evol}
\eeq
 where the
soft anomalous dimension is calculated \cite{Kulesza:2015vda,Ferroglia:2009ep} as a perturbative function in $\als$,
\beq
\pmb{\Gamma}_{ij\to klB}\left(\als\right)= \left(\frac{\als}{\pi}\right) \pmb{\Gamma}^{(1)}_{ij\to klB}+\left(\frac{\als}{\pi}\right)^2 \pmb{\Gamma}^{(2)}_{ij\to klB}+\ldots
\eeq
In order to diagonalize the one-loop soft anomalous dimension matrix we make use of the transformation~\cite{Kidonakis:1998bk}:
\begin{equation}
\pmb{\Gamma}^{(1)}_{R}  =  \mathbf{R}^{-1}\pmb{\Gamma}^{(1)}_{ij\to klB} \mathbf{R}. 
\end{equation}
Correspondingly, other matrices need to be also transformed using the diagonalization matrix $\mathbf{R}$:
$\pmb{\Gamma}^{(2)}_{R} = \mathbf{R}^{-1}\, \pmb{\Gamma}^{(2)}_{ij\to klB}\, \mathbf{R}$, $
\mathbf{H}_R  =  \mathbf{R}^{-1}\, \mathbf{H}_{ij\to klB} \, \left(\mathbf{R}^{-1}\right)^{\dagger}$,
$\mathbf{\tilde S}_{R} =  \mathbf{R}^{\dagger}\, \mathbf{\tilde S}_{ij\to klB}\, \mathbf{R}. 
$

At NLL accuracy the evolution of the soft matrix $\mathbf{S}_{ij \rightarrow t\bar tB}$ is given by the one-loop anomalous dimension matrix, see e.g. \cite{Kulesza:2015vda}. By changing the colour basis to $R$-basis,  the path ordered  exponentials in Eq.~(\ref{eq:soft:evol}), considered at NLL, reduce to simple exponentials given in terms of the  eigenvalues $\lambda_I^{(1)}$ of the soft anomalous dimension matrix $\pmb{\Gamma}^{(1)}_{R}$. Together with the LO contributions to the hard and soft function, it results in the following expression for the NLL cross section in the Mellin space
\begin{eqnarray}
&\frac{d \tilde\sigh^{{\rm (NLL)}}_{ij\tosv kl B}}{dQ^2}&(N,Q^2,\{m^2\},\muf^2, \mur^2) =  \mathbf{H}^{\mathrm{(0)}}_{R,IJ}(Q^2, \{m^2\}) \, \mathbf{\tilde S}^{\mathrm{(0)}}_{R,JI} \nn\\  &&\times
\,\Delta^i(N+1, Q^2,\muf^2, \mur^2) \Delta^j(N+1, Q^2,\muf^2, \mur^2) \nonumber \\
&&\times \, \exp\left[\frac{\log(1-2\lambda)}{2 \pi b_0}
\left(\left( \lambda^{(1)} _{J}\right)^{*}+\lambda^{(1)} _{I}\right)\right] ,
\label{eq:res:fact_diag_NLL}
\end{eqnarray}
where the color indices $I$ and $J$ are implicitly summed over, $b_0$ is the first coefficient of expansion $\beta_{\mathrm{QCD}}$ in $\als$ and $\lambda=\als b_0 \log(N)$. The trace of the product of two matrices  $\mathbf{H_R}^{\mathrm{(0)}}$ and $\mathbf{\tilde S}^{\mathrm{(0)}}_{R}$ returns the LO cross section. The incoming parton radiative factors $\Delta_i$ are now considered only at NLL accuracy.

In order to improve the accuracy of the numerical approximation provided by the NLL resummation, it is customary to include terms up to $\cal{O}(\als)$ in the expansion of the hard and soft function  leading to 
\begin{eqnarray}
&\frac{d \tilde\sigh^{{\rm (NLL\ w\ {\cal C})}}_{ij\tosv kl B}}{dQ^2}&(N,Q^2,\{m^2\},\muf^2, \mur^2) =  \mathbf{H}_{R,IJ}(Q^2, \{m^2\},\muf^2, \mur^2) \nn\\  &&\times
 \mathbf{\tilde S}_{R,JI}(Q^2,\{m^2\})  
 \Delta^i(N+1, Q^2,\muf^2, \mur^2) \Delta^j(N+1, Q^2,\muf^2, \mur^2) \nonumber \\
&&\times \, 
\exp\left[\frac{\log(1-2\lambda)}{2 \pi b_0}
\left(\left( \lambda^{(1)} _{J}\right)^{*}+\lambda^{(1)} _{I}\right)\right] ,\nn
\label{eq:res:fact_diag_NLLwC}
\end{eqnarray}
where
\beq
\mathbf{H}_{R}\, \mathbf{\tilde S}_{R} = \mathbf{H}^{\mathrm{(0)}}_{R} \mathbf{\tilde S}^{\mathrm{(0)}}_{R} + \frac{\als}{\pi}\left[\mathbf{H}^{\mathrm{(1)}}_{R} \mathbf{\tilde S}^{\mathrm{(0)}}_{R}+ \mathbf{H}^{\mathrm{(0)}}_{R} \mathbf{\tilde S}^{\mathrm{(1)}}_{R} \right] . \nn
\eeq
We will refer to this result as "NLL  w ${\cal C}$".

We compute the inclusive total cross section by integrating the expression over $Q^2$. For the differential distributions of an observable $\cal O$, in addition to the integration over $Q^2$, a function $\cal{F}_{\cal{O}}$ is introduced which includes a phase space restriction defining the observable $\cal O$:
\begin{eqnarray}
\label{eq:res:fact_diff}
&&\frac{d\tilde \sigh^{{\rm (NNLL)}}_{ij\tosv kl B}}{d\cal{O}}(N, {\cal O},\{m^2\}, \muf^2,\mur^2)  \nn
\\ 
&&= \int dQ^2\int d\Phi_3 \, {\mathrm{Tr}}\left[ \mathbf{H} (Q^2, \Phi_3, \{m^2\},\muf^2, \mur^2)\; \mathbf{S} (N+1, Q^2, \Phi_3, \{m^2\},\mur^2)  \right] \nonumber \\
&&\times\,\Delta^i(N+1, Q^2,\muf^2, \mur^2 ) \Delta^j(N+1, Q^2,\muf^2, \mur^2 ) \, {\cal F_{\cal{O}}}\left(Q^2, \Phi_3, \{m^2 \}\right)   \;.
\end{eqnarray}

The electroweak effects are included additively by matching the resummed QCD calculation to the  cross sections calculated at the complete NLO QCD and EW accuracy~\cite{Frederix:2018nkq}, indicated by NLO (QCD+EW). 
More specifically,  at the LO accuracy, apart from the  ${\cal O}(\als^2 \alpha)$ contributions, also the ${\cal O}(\als \alpha^2)$ and ${\cal O}(\alpha^3)$ terms are included. The complete NLO(QCD+EW) result, besides the ${\cal O}(\als^3 \alpha)$ correction, contains also the ${\cal O}(\als^2 \alpha^2)$, ${\cal O}(\als \alpha^3)$ and ${\cal O}(\alpha^4)$ corrections as well as the above-mentioned LO terms.

\section{Numerical predictions}

The numerical results were obtained using the same set up for input parameters as the one used in the HXSWG Yellow Report 4~\cite{deFlorian:2016spz}, i.e. \ $m_t=172.5$\,GeV, $m_H=125$ GeV, $m_W=80.385$\,GeV, $m_Z=91.188$\,GeV, $G_F=1.1663787 \times 10^{-5}$\, GeV$^{-2}$ and the  LUXqed17\_plus\_PDF4LHC15\_nnlo\_100 distribution function sets \cite{Butterworth:2015oua,Dulat:2015mca,Harland-Lang:2014zoa,Ball:2014uwa,Gao:2013bia,Carrazza:2015aoa,Manohar:2016nzj,Manohar:2017eqh}  with the corresponding values of $\als$.  The values of the NLO cross sections are obtained using the {\tt MadGraph5\_aMC@NLO} code~\cite{Alwall:2014hca,Frederix:2018nkq}, from where we also extract the QCD one loop virtual corrections needed for the hard colour matrix $\mathbf{H}^{(1)}$. All numerical results for resummed quantities were calculated  and cross-checked with two independent in-house Monte Carlo codes.

\begin{figure}[h]
\centering
\includegraphics[width=0.45\textwidth]{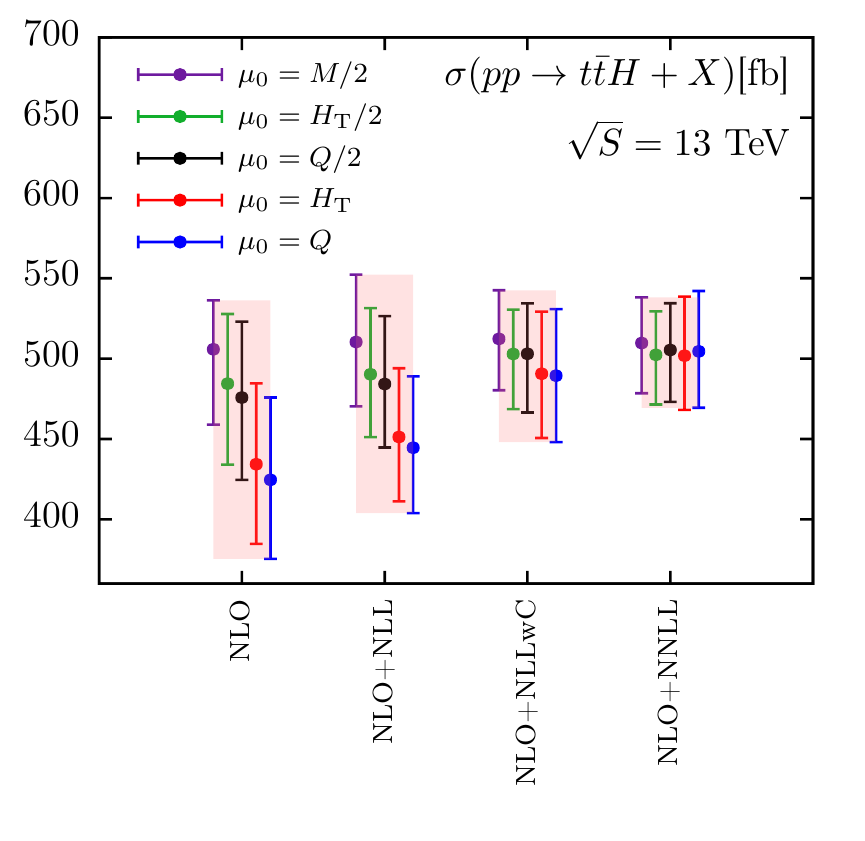}\\
\includegraphics[width=0.45\textwidth]{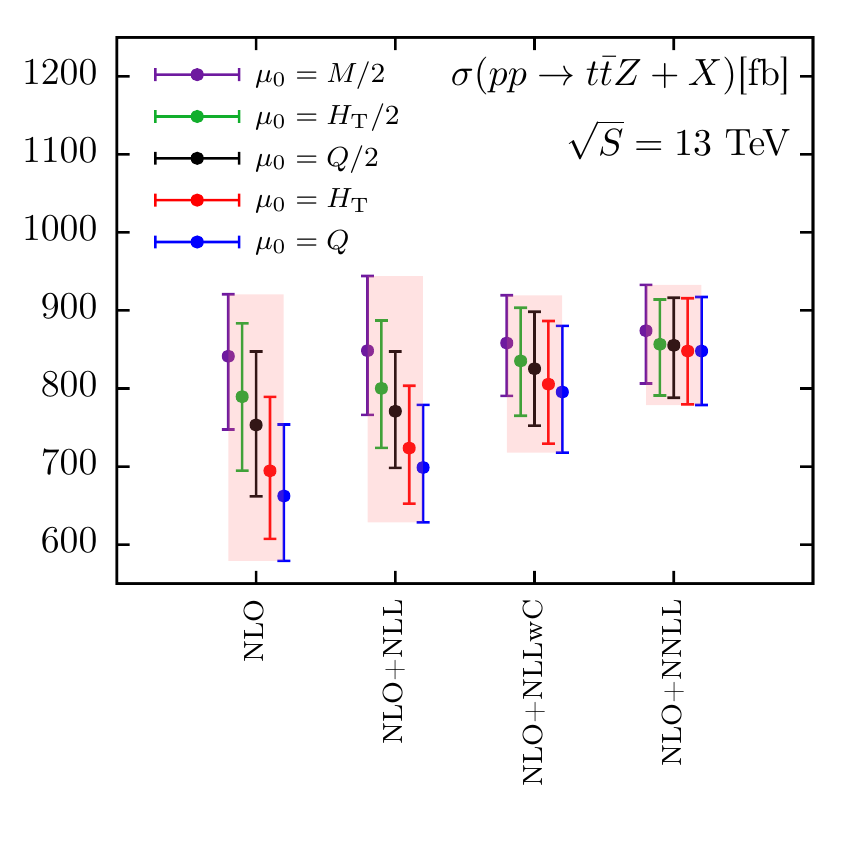}
\includegraphics[width=0.45\textwidth]{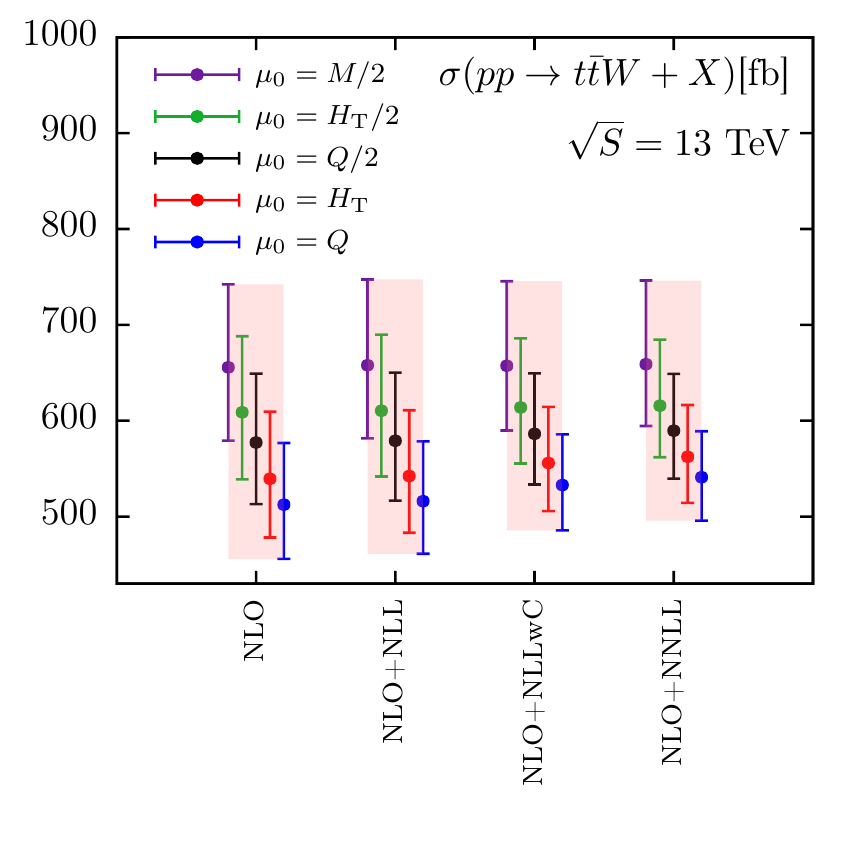}
\caption{ $pp\to \ttb$ ($B=H,Z,W$) total cross sections at various levels at theoretical accuracy.} 
\label{fig:totalxsec_13TeV_EW}
\vspace{-0.1cm}
\end{figure}

The predictions for total cross sections at various levels at theoretical accuracy for all three processes of associated top-pair production $pp\to \ttb$, $B=H,Z,W$, are shown in Fig.~\ref{fig:totalxsec_13TeV_EW}. We calculate the predictions for five choices  of the central value of the renormalization and factorization scales: $\mu_0=\mufo=\muro=Q$,  $\mu_0=\mufo=\muro=H_T$ $\mu_0=\mufo=\muro=M/2=m_t+m_B/2$ and the `in-between' values of $\mu_0=\mufo=\muro=Q/2$ $\mu_0=\mufo=\muro=H_T/2$.  The theoretical error due to scale variation is calculated using the so called 7-point method, where the minimum and maximum values obtained with $(\muf/\mu_{0}, \mur/\mu_{0}) = (0.5,0.5), (0.5,1), (1,0.5), (1,1), (1,2), (2,1), (2,2)$ are considered. 

Although the NLO(QCD+EW) results for various scale choices span quite a large range of values, we observe the results get closer  as the accuracy of resummation improves from NLL to NNLL, indicating the importance of resummed calculations.  Another manifestation of the same effect originating from soft gluon corrections is the decrease in the scale uncertainties calculated for each specific scale choice which is also progressing with increasing precision of the theoretical predictions. These trends are much stronger for $\ttbh$ and $\ttbz$ production than for $\ttbw$ due to the $gg$ channel contributing to the LO and, correspondingly, to the resummed cross section.  Given the conspicuous stability of the NLO(QCD+EW)+NNLL results, we are encouraged to combine our results obtained for various scale choices. For this purpose we adopt the method proposed by the Higgs Cross Section Working Group~\cite{Dittmaier:2011ti}. In this way, we obtain  at 13 TeV 
\begin{align}
 \sigma_{t \bar t H}^{\rm NLO+NNLL}&=504 ^{+7.6 \% +2.4 \%}_{-7.1 \% -2.4 \%} \ {\rm fb} \,, \\
 \sigma_{t \bar t Z}^{\rm NLO+NNLL}&=859 ^{+8.6 \% +2.3 \%}_{-9.5 \% -2.3 \%} \ {\rm fb} \,, \\
 \sigma_{t \bar t W}^{\rm NLO+NNLL}&=592 ^{+26.1 \% +2.1 \%}_{-16.2 \% -2.1 \%} \ {\rm fb},
\end{align}
where the first error is the scale uncertainty while the second one is the PDF uncertainty of the NLO(QCD+EW) prediction. Comparing the theoretical error  for the $\ttbz$ cross section listed above with the CMS measurement $\sigma(\ttbz)= 0.95 \pm 0.05$ (stat) $\pm 0.06$ (syst) pb~\cite{CMS:2019too}, it is clear that NNLL resummation brings the accuracy of the theoretical predictions to a level comparable with experimental precision.

\begin{figure}[t!]
\centering
\includegraphics[width=0.45\textwidth]{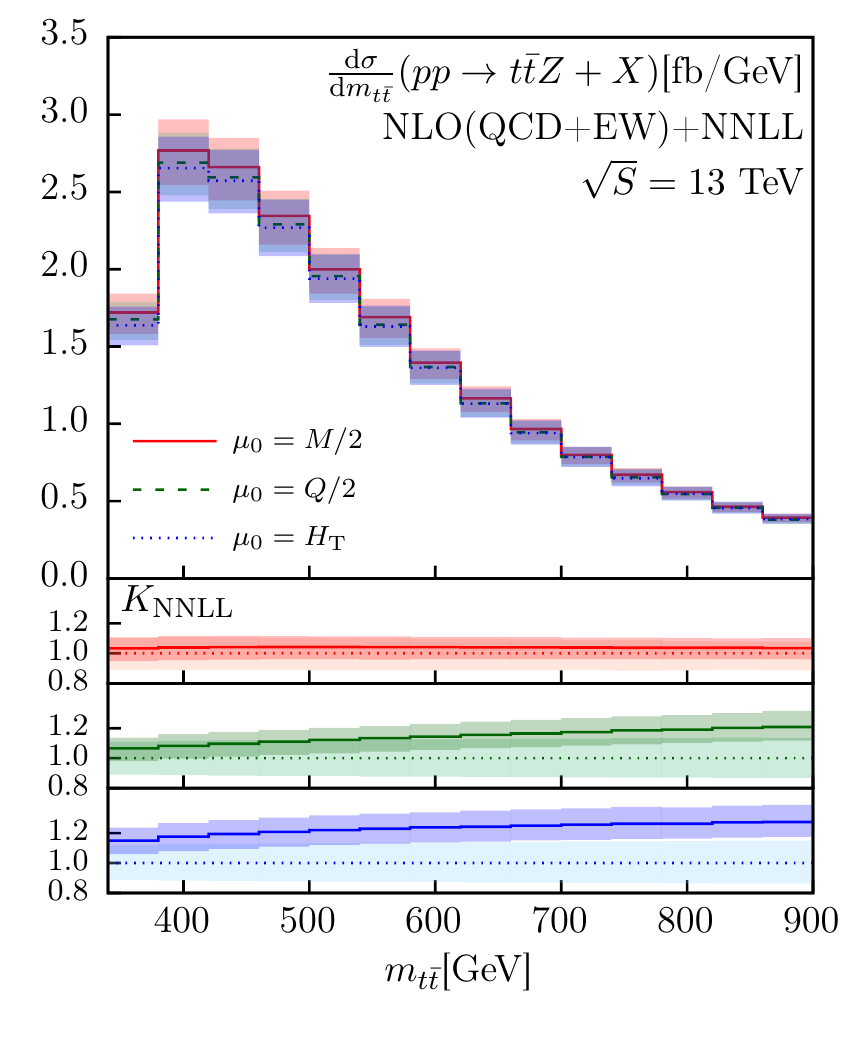}
\includegraphics[width=0.45\textwidth]{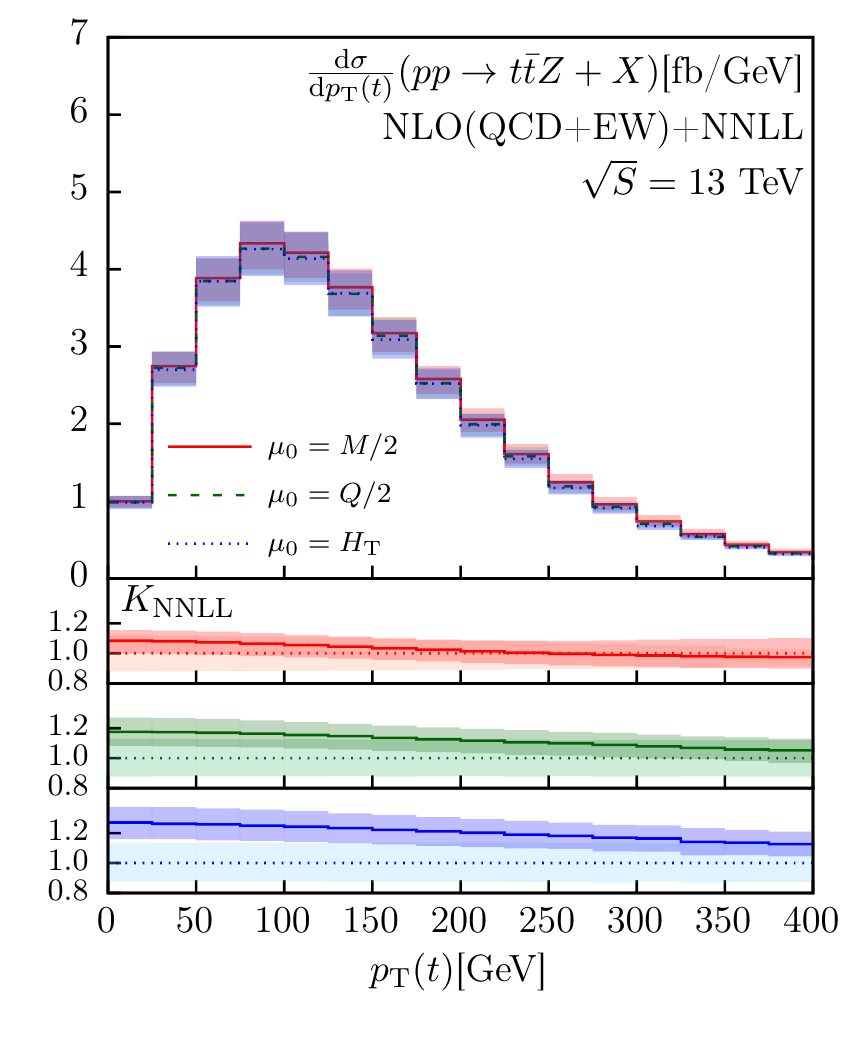}
\vspace*{-.1cm}
\caption{Predictions for $pp \to \ttbz$ differential cross sections in $m_{t \bar t}$ and $p_T(t)$. Lower panels show ratio of the NLO(QCD+EW)+NNLL and NLO(QCD+EW) distributions for three central scale choices $\mu_0=M/2$, $\mu_0=Q/2$ and $\mu_0=H_T$. Only scale uncertainties are shown.} 
\label{f:Qmttbardiff_ttZ}
\vspace*{-.1cm}
\end{figure}

\begin{figure}[h!]
\centering
\includegraphics[width=0.45\textwidth]{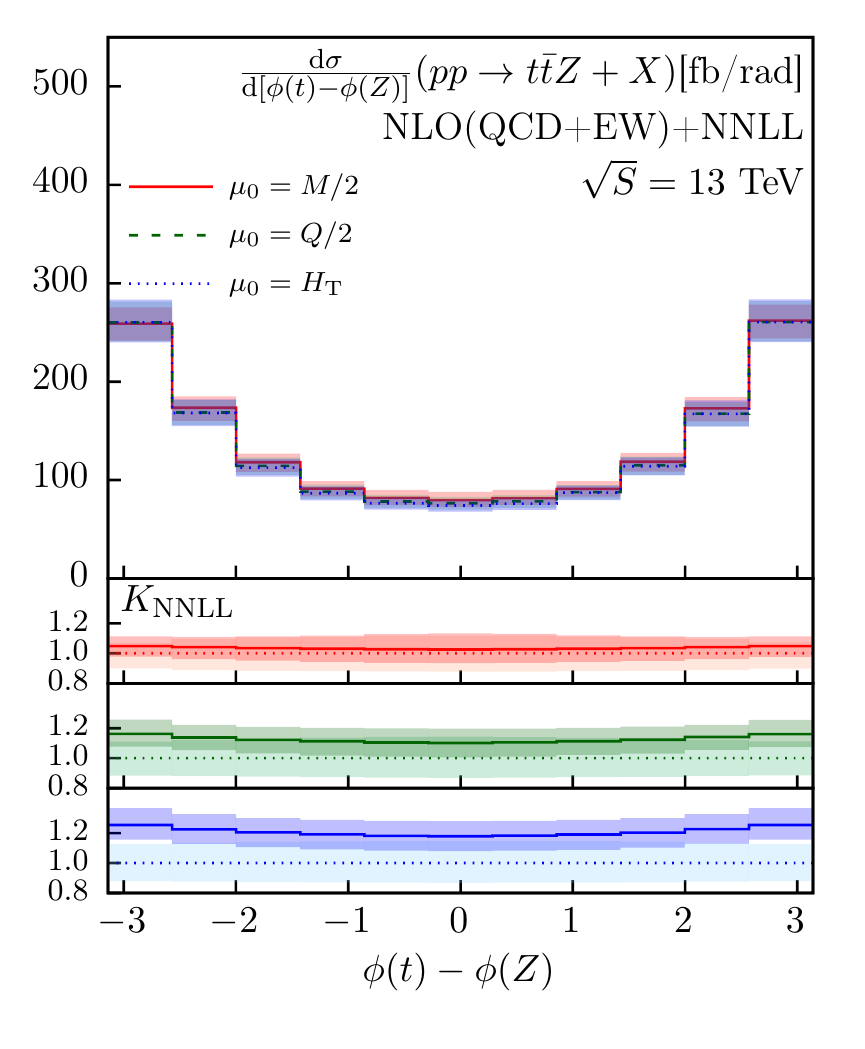}
\includegraphics[width=0.45\textwidth]{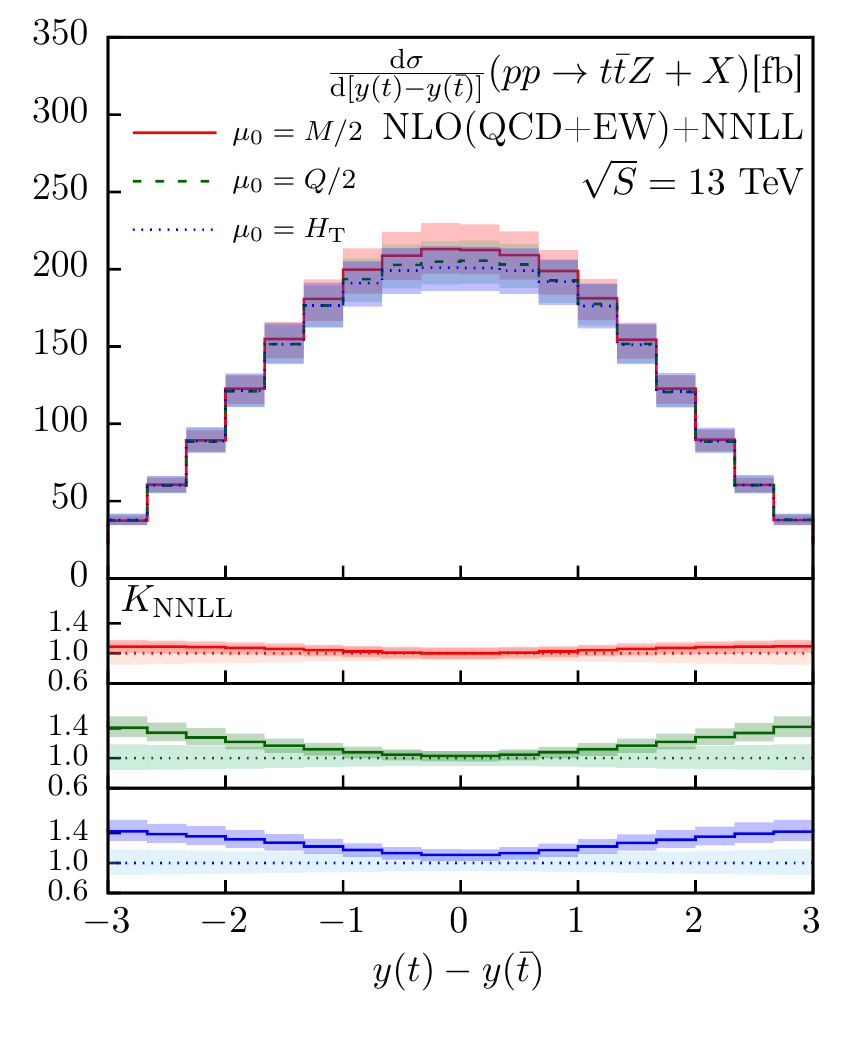}
\caption{Same as in Fig.~\ref{f:Qmttbardiff_ttZ} but for differential cross sections  in  $\phi(t)-\phi(Z)$  and in $y(t)-y(\bar t)$.} 
\label{f:phidiff_ttZ}
\vspace*{-.1cm}
\end{figure}

\begin{figure}[h!]
\centering
\includegraphics[width=0.45\textwidth]{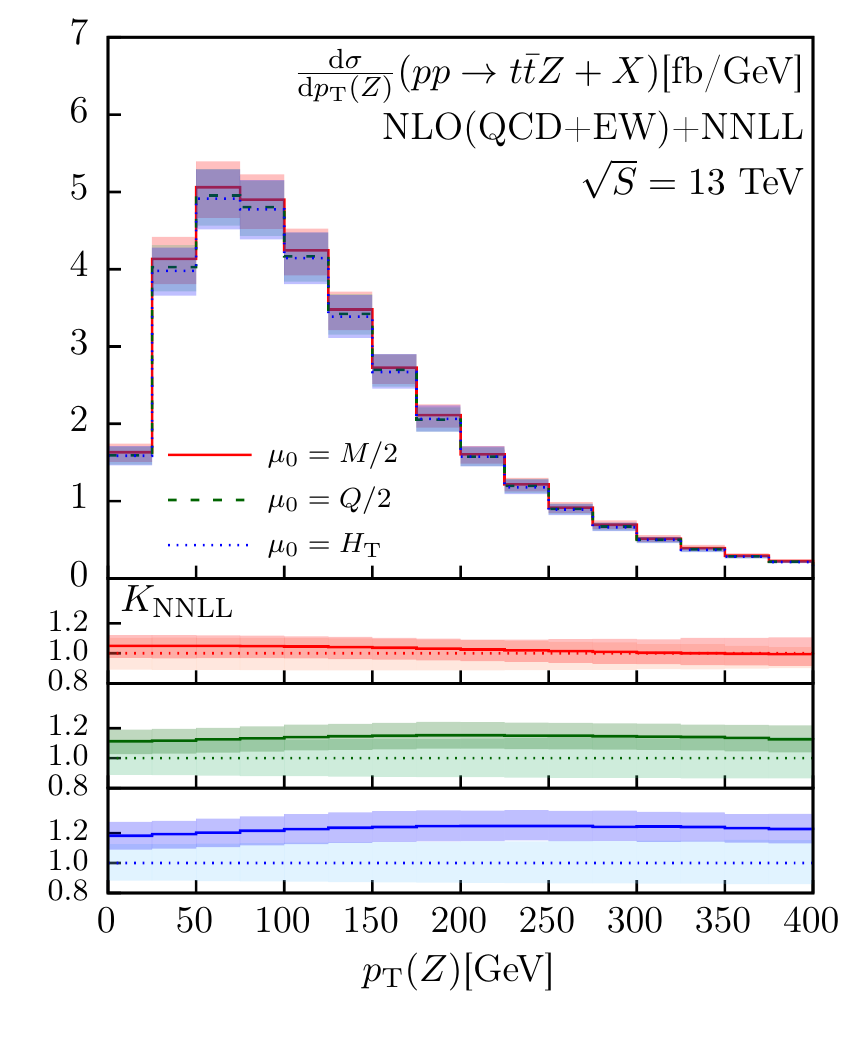}
\includegraphics[width=0.45\textwidth]{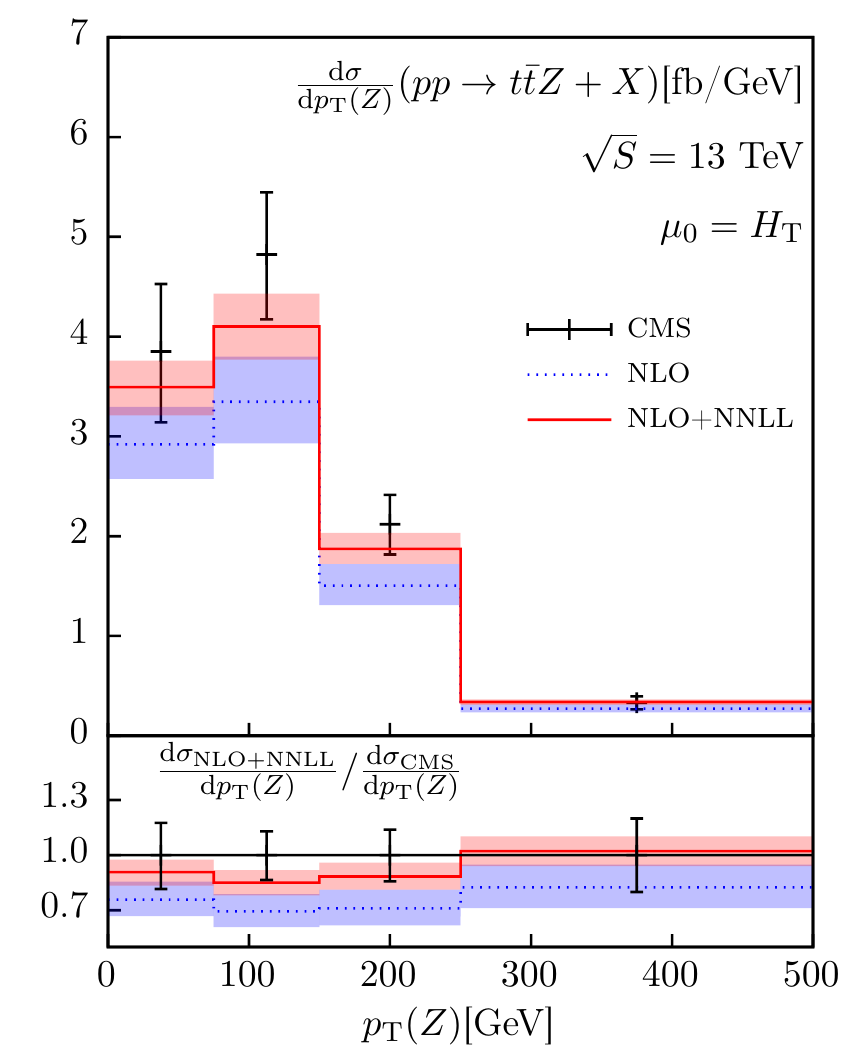}
\caption{Transverse momentum distribution $p_T(Z)$ of the $Z$ boson produced in the $pp \to \ttbz$ process. Left: Same as in Fig.~\ref{f:Qmttbardiff_ttZ} but for  $p_T(Z)$. Right: comparison of the NLO(QCD+EW) and  NLO(QCD+EW)+NNLL predictions calculated at $\mu_0=H_T$ with the CMS data~\cite{CMS:2019too}.} 
\label{f:pTdiff_ttZ}
\end{figure}

As discussed above, the presented formalism allows to study a number of differential distributions. In particular, we have access to observables that are invariant under boosts from the hadronic center-of-mass frame to the partonic center-of-mass frame. In Figs.~\ref{f:Qmttbardiff_ttZ}--\ref{f:phidiff_ttZ} we show selected NLO(QCD+EW) + NNLL distributions for the process with the highest cross section,  $pp \to \ttbz$, for three representative scale choices: $\mu_0=M/2$, $\mu_0=Q/2$ and $\mu_0=H_T$.   We refer the reader to~\cite{Kulesza:2020nfh} for differential cross sections for the $\ttbh$, $\ttbw$ production as well as  additional distributions for the $\ttbz$ process. The top panels of Figs.~\ref{f:Qmttbardiff_ttZ}, \ref{f:phidiff_ttZ}  and \ref{f:pTdiff_ttZ} (left) demonstrate an excellent agreement for the  NLO(QCD+EW)+NNLL predictions obtained for the three scale choices. The lower three panels in the figures show ratios of the NLO(QCD+EW)+NNLL distributions to the NLO(QCD+EW) distributions, calculated for different values of $\mu_0$. The dark shaded areas indicate the scale errors of the  NLO(QCD+EW)+NNLL predictions, while light-shaded areas correspond to  the scale errors of the  NLO(QCD+EW) results. We observe that the ratios can differ substantially depending on the final state, observable or the central scale. Generally, the NNLL resummation has the biggest impact on the predictions obtained for $\mu_0=H_T$ among the three scale choice we study. The ratios show that resummation can contribute as much as ca. 30\%, up to 40\%, correction to the $\ttbz$ distribution at this scale choice.  Fig. \ref{f:pTdiff_ttZ} focuses on the $p_T (Z)$ distribution: on the left side we show the NLO(QCD+EW) + NNLL distributions for the three different scale choices like in Figs.~\ref{f:Qmttbardiff_ttZ}, \ref{f:phidiff_ttZ}, while the right plot shows a  comparison of the CMS data~\cite{CMS:2019too} to the NLO(QCD+EW) and our NLO(QCD+EW)+NNLL predictions for the scale choice $\mu_0=H_T$. From the figure, it is clear that the resummed NNLL corrections bring the theoretical predictions closer to data and lead to a significant reduction of the scale dependence error.\\
{\bf Acknowledgements:} This work has been supported by the DFG grant KU3103/2 and the National Science Center grants No. 2017/27/B/ST2/02755 and 2019/32/C/ST2/00202.

\vspace*{-.5cm}

\end{document}